\documentclass[twocolumn,showpacs,preprintnumbers,amsmath,amssymb]{revtex4}
\usepackage[english]{babel}
\usepackage[T1]{fontenc}
\usepackage[cp1250]{inputenc}
\usepackage{graphicx}
\usepackage{dcolumn}
\usepackage{bm}
\usepackage{ulem}
\newcommand{\vc}[1]{\mathbf{#1}}

\begin{document}

\title{Non-destructive study of non-equilibrium states of cold, trapped atoms}

\author{Maria Brzozowska}
\author{Tomasz M. Brzozowski}
\author{Jerzy Zachorowski}
\author{Wojciech Gawlik}
\email{gawlik@uj.edu.pl} \affiliation{Marian Smoluchowski Institute of Physics,
Jagiellonian University, Reymonta 4, PL 30-059 Cracow}
\homepage{http://www.if.uj.edu.pl/ZF/qnog/}
\date{\today}

\begin{abstract}
Highly sensitive, non-destructive, real-time spectroscopic determination of the
2D kinetic momentum distribution of a cold-atom sample is performed with the
three-beam measurement of the recoil-induced resonances. The measurements
performed with an operating magneto-optical trap reveal slow velocity drifts
within a stationary atomic cloud and strong anisotropy and asymmetry of the
non-Maxwellian momentum distribution. The developed method can be easily
extended to 3D.

\end{abstract}


\maketitle

Most experiments with cold, dilute atomic gases employ magneto-optical traps
(MOT), which yield temperatures in a range of hundreds to a few $\mu$K. Further
traps and cooling stages can be applied for reaching the quantum degeneracy
regime. This requires matching of the momentum distributions of various traps.
Knowledge of such distributions is also essential for quantum state diagnostics
of the trapped sample. Below, we present reliable method of 2D momentum
diagnostics based on the so called recoil-induced resonances (RIRs) and apply
it to the detailed study of non-standard momentum distributions in a MOT.

The first unambiguous observation of RIRs was made in a 1D optical
lattice~\cite{gry94} filled with atoms much colder than in a standard MOT. RIR
signals were also seen in optical molasses~\cite{mea94,fisch01}, in a cold
atomic beam~\cite{dom01} and with atoms released from a MOT~\cite{chen01}. The
influence of the recoil effect on the probe absorption and four wave mixing
spectra has been recently demonstrated in a continuously working MOT,
i.e. with all light and magnetic fields on, in~\cite{brz05}.

In this Letter we present evidence of three different kinds of anisotropy of
the momentum distribution in an operating MOT. The measurements were
conducted using our three-beam, RIR-based method developed for
simultaneous probing of the momentum distribution in two perpendicular
directions. The method extends the principle of 1D thermometry as suggested
in Ref.~\cite{mea94}. Important feature of our extension is that 2D
information is acquired simultaneously in one measurement. The method can be
extended to 3D.

\begin{figure}[h]
\includegraphics[scale=0.65]{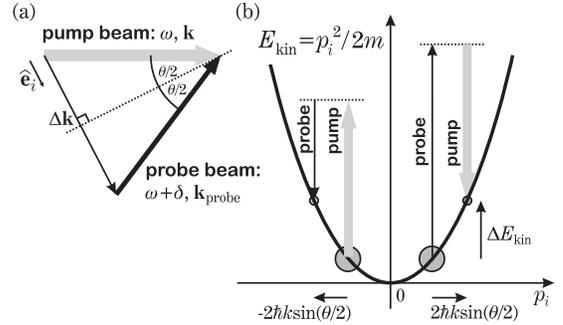}
\caption{\label{fig01}Recoil-induced resonances: (a) geometry of the laser
beams, (b) atomic kinetic energy and momentum changes due to the Raman process
coupling two kinetic states. Circles symbolize populations $\Pi(\vc{p})$ of
these states.}
\end{figure}

The RIRs result from a stimulated Raman process, which couples two kinetic
states of free moving atoms (Fig.~\ref{fig01}). Two laser beams, the pump and
the probe, of frequencies $\omega$ and $\omega+\delta$, respectively, drive the
Raman transition after which the atoms gain kinetic energy $\Delta
E_{\mathrm{kin}}=\hbar \delta$ and change momentum $\vc{p}$ by $\Delta
\vc{p}=\hbar\Delta\vc{k}=\pm2\hbar k \hat{\vc{e}}_i \sin\theta/2$, where $k$ is
the modulus of the light wave vector, $\theta$ is the angle between the beams,
and $\hat{\vc{e}}_i$ is the unit vector perpendicular to the bisector of
$\theta$. The non-zero amplitude of the considered Raman resonance arises from
different populations of the given kinetic states. When recorded in absorption,
the RIR shape is proportional to a derivative of the momentum distribution
$\partial\Pi(p_i)/\partial p_i$, where
$p_i=\vc{p}\cdot\hat{\vc{e}}_i$~\cite{gry94,ver96,guo92,brz05}. This direct
relation of the RIR signal to $\Pi(\vc{p})$ allows convenient and accurate
measurement of the kinetic momentum distribution in a cold atomic sample,
provided that the distribution is sufficiently narrow.

Important advantage of the RIR method is its directional selectivity.
$\Pi(\vc{p})$ is probed in a given direction, specified by $\Delta{\vc{p}}$,
i.e. by angle $\theta$ (Fig.\ref{fig01}a). Hence, apart from applications to
standard 1D velocimetry \cite{mea94,dom01}, RIRs can also be used for studies
of a possible momentum distribution anisotropy in non-equilibrium states of a
cold atomic sample.

\begin{figure}[ht!]
\includegraphics[scale=0.65]{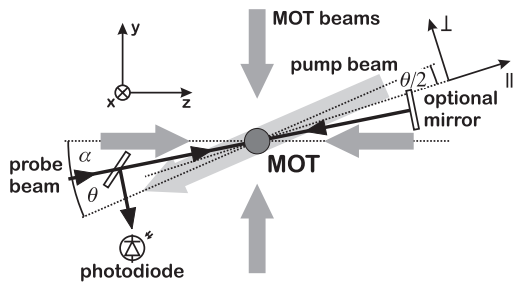}
\caption{\label{fig02}The layout of the experiment. The pump and probe beams
intersect in a cloud of cold atoms. With the optional mirror we realize a
three-beam configuration in which the momentum exchange is allowed in two
directions: $\perp$ and ${||}$. Third pair of the MOT beams along $x$ and the
MOT coils are not shown in this figure.}
\end{figure}

Our experiment (Fig.~\ref{fig02}) employs a standard vapor-loaded
MOT~\cite{raab87} with $^{85}$Rb atoms. Two extra beams intersect in the trap
center: the pump and the probe. The probe beam is directed at a small angle
$\alpha=3^{\circ}$ to the MOT beams (propagating along $z$), and the pump is at
$\theta=5^{\circ}$ to the probe. The probe beam can be detected either directly
or after retroreflection. The setup with retroreflected probe enables the
measurement of $\Pi(\vc{p})$ simultaneously along two perpendicular directions:
$\perp$, for angle $\theta$ between the pump and the nearly co-propagating
probe, and ${||}$, for angle $180^\circ-\theta$ between the pump and the nearly
counter-propagating probe. When $\alpha$ and $\theta$ are sufficiently small,
$\perp$ and ${||}$ almost coincide with the $y$ and $z$ directions,
respectively. Both pump and probe beams are derived from diode lasers
synchronized by injection-locking and are blue-detuned from the trapping
transition $^2$S$_{1/2}(F=3)$--$^2$P$_{3/2}(F'=4)$ by $\Delta=2\pi\cdot$140 MHz
$\approx 23.3\,\Gamma$, where $\Gamma$ denotes the natural linewidth. Such big
detuning reduces the perturbation of atoms to a very low level (scattering rate
$\propto1/\Delta^2$) which is essential for non-destructive measurements.
Moreover, non-resonant pump eliminates overlap of the RIRs and Raman-Zeeman
resonances~\cite{brz05} hence facilitates interpretation of the results.
Despite large $\Delta$, the pump and probe beams drive the Raman signal with a
sufficiently large amplitude and signal-to-noise ratio for the pump beam
intensities 5-35~mW/cm$^2$.

The probe beam is scanned by $\delta\approx\pm 1$~MHz around frequency $\omega$
of the pump. The probe and pump photons induce Raman transitions between atomic
kinetic states separated by $\pm\hbar\delta$. Since the polarization of the
pump and probe beams is chosen to be the same, the atoms undergo Raman
transitions with $\Delta m_F=0$. Hence, the internal atomic state does not
change and the only states that have to be considered are the external states
associated with the kinetic energy of the atomic center-of-mass. The
multi-level structure of $^{85}$Rb can thus be reduced to a set of independent
two-level systems which allows straightforward application of the basic RIR
theory \cite{gry94,brz05,ver96}. With the assumption that $\Pi(\vc{p})$ is the
Maxwell-Boltzmann distribution, the RIR signal $s(\delta)$~\cite{ver96,brz05}
recorded with the retroreflected probe is given by two contributions. The
narrow one results from the Raman process involving the pump and the probe beam
making small angle $\theta$ and the wide one is for angle $180^{\circ}-\theta$.
The signal is
\begin{equation} \label{eq:signal} s(\delta)\propto
-A_{\perp}\delta \exp\left(-\frac{\delta^2}{\xi_{\perp}^2}\right)
-A_{||}\delta
\exp\left(-\frac{(\delta-\delta_0)^2}{\xi_{||}^2}\right),
\end{equation}
where, for small $\theta$,
$\xi_{\perp}^2\approx2k_Bk^2m^{-1}\theta^2\tau_{\perp}$ and
$\xi_{||}^2\approx8k_Bk^2m^{-1}\tau_{||}$. $\tau_{\perp}$ and
$\tau_{||}$ are the distribution widths in the $\perp$ and $||$
directions in the temperature units, $A_{||}$ and $A_{\perp}$ are
the amplitudes of the corresponding contributions, $m$ is the atomic
mass, $k_B$ is the Boltzmann constant, and $\delta_0$ is the
possible frequency shift between the $\perp$ and $||$ contributions,
to be discussed later.

\begin{figure}[ht]
\includegraphics[scale=0.4]{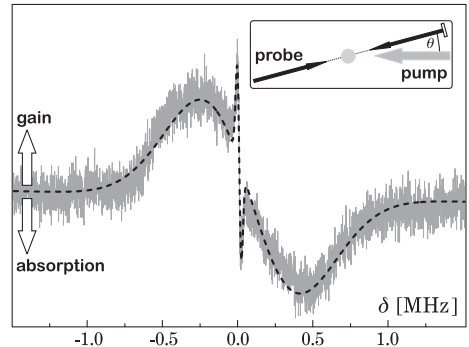}
\caption{\label{fig03}Transmission spectrum of a retroreflected
probe (gray) and the theoretical prediction (dashed) of Eq.(1).
Inset: the beam setup. The MOT beams have intensity
$I_{\mathrm{MOT}}=13.8\,\, \mathrm{mW/cm}^2$ per beam and are
detuned from the trapping transition by
$\Delta_{\mathrm{MOT}}=-3\,\Gamma$, the repumper beam intensity
$I_\mathrm{REP}=15\,\,\mathrm{mW/cm}^2$, the axial magnetic field
gradient $\partial_x B=12\,\, \mathrm{Gauss/cm}$, the pump beam
intensity $I=33\,\,\mathrm{mW/cm}^2$, and the probe beam intensity
0.3~mW/cm$^{2}$.}
\end{figure}

Typical example of the retroreflected-probe transmission spectrum is shown in
Fig.~\ref{fig03}. It exhibits two distinct resonant contributions, predicted by
eq.~\eqref{eq:signal}. The wide contribution is shifted with respect to the
narrow one by 72.4 kHz, which indicates a 2.8-cm/s average velocity component
in the ${||}$ direction. We thus observe an atomic drift within a cloud, which
as a whole remains stationary. We understand this as a dynamic effect resulting
from a small difference of the radiation pressures intrinsic to a MOT with
retroreflected trapping beams. Indeed, the observed shift increases when the
imbalance is purposely increased. Strong imbalance normally produces a
displacement of the atomic-cloud center of mass. In our case this displacement
is too weak to be detected by standard imaging technique, while the anisotropic
atomic flow, even one order of magnitude slower than the mean thermal velocity,
is well measurable with our method.

As the velocity distributions derived from the signal in Fig.~\ref{fig03} are
Gaussian, one can determine the values $\tau_\perp=172\pm6\,\mu$K and
$\tau_{||}=170\pm3\,\mu$K. The equality of these $\tau$s implies
thermodynamical equilibrium and allows their interpretation as temperature $T$,
despite the slow drift. The equilibrium persists for various MOT-light
intensities due to the fact that total intensities of each pair of the MOT
beams remain the same. The observed nearly linear increase of $T$ with the
total MOT-beam intensity agrees well with previous
reports~\cite{Lett89,Wallace94,Ye2002}.

\begin{figure}[ht]
\includegraphics[scale=0.4]{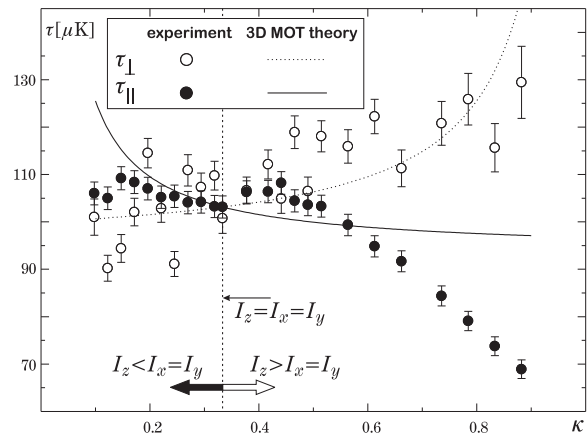}
\caption{\label{fig04} Widths of the kinetic momentum distributions
measured as a function of relative intensity $\kappa$ in the two
perpendicular directions, $\tau_\perp$ (hollow circles)and
$\tau_{||}$ (filled circles). For the equilibrium ($\kappa=1/3$),
$I_\mathrm{MOT}=6.8\,\,\mathrm{mW/cm}^2$, $\tau_\perp=100\pm
3\,\,\mathrm{\mu K}$, and $\tau_{||}=103\pm2\,\,\mathrm{\mu K}$.
Other parameters as in Fig.~\ref{fig03}. Theoretical curves are
plotted according to Refs.~\cite{gajda94,gajda05}.}
\end{figure}

The thermodynamics of the system becomes highly non-trivial when the trapping
light is unevenly distributed between the MOT beam pairs. It was predicted that
for such conditions the width of kinetic momentum distribution shows
directional dependence \cite{gajda94}. Using the simultaneous measurement of
the momentum width in two perpendicular directions, we attempted to observe
such anisotropic non-equilibrium state of the cold-atom cloud. For this reason,
we changed intensity balance between the longitudinal ($I_z$) and transverse
($I_x$, $I_y$) MOT beam pairs, while keeping the total intensity
$I_0=I_x+I_y+I_z$ constant. We define parameter $\kappa$ as the relative
intensity of $I_z$, $I_z=\kappa I_0$, $I_x=I_y=(1-\kappa)I_0/2$. The results of
the measurement of $\tau_{||}$ and $\tau_{\perp}$ for different values of
$\kappa$ are depicted in Fig~\ref{fig04}. For equal partition of the trapping
intensity ($\kappa=1/3$), the widths of kinetic momentum distributions are the
same, as expected. However, when $\kappa$ increases, $\tau_{||}$ and
$\tau_{\perp}$ follow opposite trends, which is evidence of kinetic momentum
anisotropy in a MOT working MOT and thereby its non-equilibrium state. A
similar anisotropy was recently observed also in an optically dense sample
\cite{vorozcovs}. We notice that $\tau_{\parallel}+2\tau_{\perp}$, which is the
measure of $v_{\parallel}^{2}+2v_{\perp}^{2}$, is constant within $\pm2\%$ over
the whole measured range of $\kappa$. The decrease of $\tau_{||}$ with the
growing $\kappa$ is due to the fact that the heating associated with
spontaneous emission is isotropic, whereas the cooling rate is higher for the
direction with the increased intensity. The momentum anisotropy becomes
manifest because the density of the atoms is too small to provide efficient
thermalization. Indeed, simple estimation for typical conditions and Rb-Rb
collision cross-section
$\sigma_{\rm{Rb-Rb}}=3\cdot10^{-13}$~cm$^2$~\cite{rapol01} yields the atomic
collision rate below 1~Hz in our trap, while the friction coefficient, in
frequency units, is in the kHz range.

The theoretical behavior of $\tau_{||}$ and $\tau_{\perp}$ according to
Refs.~\cite{gajda94,gajda05} is plotted in Fig.~\ref{fig04} along with the
experimental data. They exhibit similar qualitative dependence (the decrease of
$\tau_{||}$ and the increase of $\tau_{\perp}$ with growing $\kappa$), but the
existing theory fails to reproduce the exact shape of the experimental
dependence. This discrepancy is due to additional mechanism of sub-Doppler
cooling, not included in the calculations of Ref.~\cite{gajda94}. Indeed, the
increase of $I_z$ accompanied by attenuation of $I_x$ and $I_y$ results in
efficient quasi-1D cooling scheme in the $\sigma^{+}-\sigma^{-}$ optical
molasses~\cite{dali89}. Evidence of this cooling is provided by the values of
$\tau_{||}$ falling to 70~$\mu$K, well below the Doppler cooling limit of
140~$\mu$K. Importance of sub-Doppler cooling for anisotropy of momentum
distribution in cold atomic samples has been previously noted in optical
molasses~\cite{Jav}.

In the situation discussed above, the MOT beams were carefully aligned which
resulted in high stability of the trapped-atom cloud, even for the largest
departures from the equal partition of the trapping light intensity, and
allowed fitting of the RIR signals by eq.~\eqref{eq:signal}. The sole
manifestation of the non-equilibrium of the sample was the $\Pi(p_{\perp})$ vs.
$\Pi(p_{||})$ anisotropy.
\begin{figure}[h]
\includegraphics[scale=0.62]{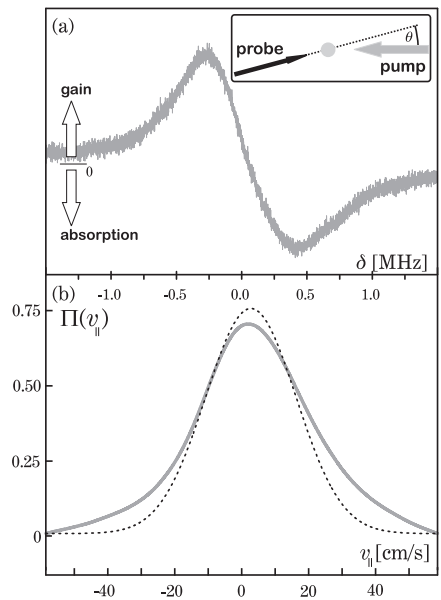}
\caption{\label{fig05}(a) The probe transmission spectrum in a
misaligned MOT for $I_{\mathrm{MOT}}=3.1\,\,\mathrm{mW/cm}^2$,
$\Delta_{\mathrm{MOT}}=-2.25 \Gamma$ and other experimental
parameters as in Fig~\ref{fig03}. As shown in inset, the probe-beam
is not retroreflected, hence only the wide RIR contribution is
present. (b) The actual velocity distribution $\Pi(v_{||})$ (grey
thick line) obtained by integrating the probe transmission spectrum
and the Gaussian reference (dotted line), defined in text.}
\end{figure}
The thermodynamical equilibrium can be altered yet in a different way, namely
by enhancing imbalance between the counter-propagating MOT-beam radiation
pressures. Fig.~\ref{fig05}a depicts the RIR recorded with a standard 1D,
two-beam arrangement applied to the case when the MOT beams were slightly
misaligned and tuned closer to resonance. The 1D thermometry was accomplished
by replacing the optional mirror in Fig.~\ref{fig02} by a photodiode. In this
configuration, the pump and the probe make angle $180^\circ-\theta$ and the
recorded signal is proportional only to $\partial \Pi(p_{||}) /\partial
p_{||}$. Its shape deviates from a derivative of a Gaussian. By integrating the
signal and scaling to the velocity units, the actual distribution of velocity
component in the $||$ direction, $v_{||}$, can be retrieved. Fig.~\ref{fig05}b
shows such a distribution obtained from the experimental signal and the
idealized Gaussian reference curve of the same area and of the width derived
from the positions of the minimum and maximum of the RIR signal in
Fig.~\ref{fig05}a. Non-standard properties of a stable atomic gas, revealed in
our experiment call for thorough theoretical analysis with proper accounting
for cooling and heating mechanisms.

In conclusion, we have developed three-beam spectroscopic method of determining
the momentum distributions of cold, trapped atoms, based on recoil-induced
resonances. The method is non-destructive, highly sensitive and provides
multi-dimensional momentum determination in a single measurement. Its potential
has been demonstrated by study of three different momentum distributions of
atoms in the operating magneto-optical trap: (i) thermodynamic equilibrium with
well defined temperature and Gaussian momentum distribution with a slow
velocity drift; (ii) non-equilibrium state characterized by Gaussian
distributions with drastically different widths in the longitudinal and
transverse directions; (iii) non-equilibrium state of non-Gaussian momentum
distribution along one direction. The result (ii) qualitatively confirms
theoretical predictions of Ref.~\cite{gajda94} and indicate need for more
refined MOT theory. Our method can be particularly useful for studies of
anisotropy in optical molasses~\cite{Jav}, 2D MOTs~\cite{Dieck98}, etc. The
described method can be also straightforwardly applied to 3D case by
introducing additional pump beam in the $xz$ plane in Fig.~\ref{fig02}. To
avoid overlapping of the RIR signals associated with all $\Delta{\vc{p}}$
directions, the frequency of the additional pump could be shifted. Such an
approach can be used for the non-destructive, on-line diagnostics of the atom
dynamics in a trap carried out independently and simultaneously with other
spectroscopic measurements. The method should also be applicable to
quantum-degenerate gases. In fact, the widely used Bragg spectroscopy is based
on the same principle of momentum and energy transfer. Measuring the Bragg-beam
transmission in our three-beam geometry, rather than imaging BEC can thus
become a valuable, non-destructive alternative.

This work was supported by the Polish Ministry of Science and
Information Society Technologies and is part of a general program on
cold-atom physics of the National Laboratory of AMO Physics in
Toruń, Poland. Authors would like to thank Mariusz Gajda for his
illuminating discussion  and Krzysztof Sacha and
Dmitry Budker for their valuable comments.

\end{document}